\documentclass[twocolumn,prl,amsmath,amssymb,floatfix,reprint,square,comma,numbers]{revtex4-1}
\usepackage{graphicx}
\usepackage{ulem}
\usepackage{amstext}
\usepackage{makeidx}
\usepackage[colorlinks=true,linkcolor=blue]{hyperref}
\usepackage{color,soul}
\usepackage{amsmath}
\usepackage{braket}

\begin{document}

\title{Observation of multi-component atomic Schr\"odinger cat states of up to 20 qubits}

\author{Chao Song$^{1}$}
\thanks{C. S., K. X., and H.K. L. contributed equally to this work.}
\author{Kai Xu$^{2,4}$}
\thanks{C. S., K. X., and H.K. L. contributed equally to this work.}
\author{Hekang Li$^{2}$}
\thanks{C. S., K. X., and H.K. L. contributed equally to this work.}
\author{Yuran Zhang$^{2,5}$, Xu Zhang$^1$, \mbox{Wuxin Liu$^1$}, Qiujiang Guo$^1$, Zhen Wang$^1$, Wenhui Ren$^1$, Jie Hao$^3$, Hui Feng$^3$}
\author{\mbox{Heng Fan$^{2,4}$}}
\email{hfan@iphy.ac.cn}
\author{Dongning Zheng$^{2,4}$}
\email{dzheng@iphy.ac.cn}
\author{Dawei Wang$^{1,4}$}
\author{H. Wang$^{1,6}$}
\email{hhwang@zju.edu.cn}
\author{Shiyao Zhu$^{1,6}$}
\affiliation{$^{1}$ Interdisciplinary Center for Quantum Information and \mbox{Zhejiang Province Key Laboratory of Quantum Technology and Device,}
\mbox{Department of Physics, Zhejiang University, Hangzhou 310027, China},
$^2$ \mbox{Institute of Physics, Chinese Academy of Sciences, Beijing 100190, China},
$^3$ \mbox{Institute of Automation, Chinese Academy of Sciences, Beijing 100190, China},
$^4$ CAS Center for Excellence in Topological Quantum Computation, University of Chinese Academy of Sciences, Beijing 100190, China,
$^5$ Beijing Computational Science Research Center, Beijing 100094, China,
$^6$ Synergetic Innovation Center of Quantum Information and Quantum Physics, University of Science and Technology of China, Hefei, Anhui 230026, China
}
\date{\today}

\begin{abstract}
We report on deterministic generation of 18-qubit \textit{genuinely entangled} Greenberger-Horne-Zeilinger (GHZ) state
and multi-component atomic Schr\"{o}dinger cat states of up to 20 qubits
on a quantum processor, which features 20 superconducting qubits interconnected by a bus resonator.
By engineering a one-axis twisting Hamiltonian enabled by the resonator-mediated interactions, 
the system of qubits initialized coherently 
evolves to an over-squeezed, non-Gaussian regime,
where atomic Schr\"{o}dinger cat states, i.e., superpositions of atomic coherent states including GHZ state, appear at specific time intervals in excellent agreement with theory. 
With high controllability, we are able to take snapshots of the dynamics by plotting
quasidistribution $Q$-functions of the 20-qubit atomic cat states, and
globally characterize the 18-qubit GHZ state which yields a fidelity of $0.525\pm0.005$ confirming \textit{genuine eighteen-partite entanglement}.
Our results demonstrate the largest entanglement controllably created so far in solid state architectures, 
and the process of generating and detecting multipartite entanglement may promise applications in 
practical quantum metrology, quantum information 
processing and quantum computation.
\end{abstract}

\maketitle

The capability of controllably entangling multiple particles 
is central to fundamental test of quantum
theory~\cite{Zurek2001}, and represents a key prerequisite for quantum information processing. 
There exist various kinds of multipartite entangled
states, among which the Greenberger-Horne-Zeilinger (GHZ) states, i.e., the 2-component atomic Schr\"{o}dinger cat states, are
particularly appealing and useful~\cite{Greenberger1990}. These states
play a key role in quantum-based technologies, including open-destination quantum
teleportation~\cite{Zhao2004}, 
concatenated error correcting codes~\cite{Knill2005}, quantum simulation~\cite{Song2018}, and high-precision spectroscopy measurement~\cite{Leibfried2004}. 
In principle, the number of particles that can be deterministically entangled in a quantum processor is a benchmark of its capability in processing quantum information.
However, it is difficult to scale up this number 
since the conventional step-by-step gate methods require long control sequences which increase exposure to perturbing noise. 
A shortcut is to realize the free evolution under a
nonlinear Hamiltonian with, e.g., one-axis twisting, and the system of qubits initialized in an atomic coherent state 
is predicted to evolve to squeezed spin states~\cite{Ma2011}, and then to the multi-component
atomic Schr\"{o}dinger cat states~\cite{Agarwal1997}, i.e., superpositions of atomic coherent states including GHZ state~\cite{Zheng2001}.

Engineering fully controllable and highly coherent multipartite quantum computing platforms 
remains an outstanding challenge. Several physical platforms are being explored~\cite{Barends2016,Wang2016,Bernien2017,Lu2017,Song2017,Friis2018},
and a series of experiments for generating multipartite entanglement 
were reported~\cite{DiCarlo2010,Neeley2010,Paik2016,Song2017,Friis2018,YWang2018,Monz2011,Zhong2018,Wang2018,Gong2019}. 
Some of these experiments involve local detections of only the subsystems~\cite{Friis2018,YWang2018}. 
Multipartite entanglement, in particular the GHZ state which possesses global entanglement, 
would be better characterized by synchronized detections of all system parties and was achieved
with 14 trapped ions~\cite{Monz2011}, 12 photons~\cite{Zhong2018}, 18 photonic qubits exploiting 
6 photons~\cite{Wang2018}, and 12 superconducting qubits~\cite{Gong2019}. 
In particular, previously we reported the production and full tomography of the 10-qubit GHZ state
and the implementation of high-fidelity two-qubit gate
with an all-to-all connected superconducting quantum processor~\cite{Song2017,Guo2018}, where each
qubit can be individually controlled and qubit-qubit interactions can be turned on and off as desired.

\begin{figure*}[t]
\includegraphics[clip=True,width=140mm]{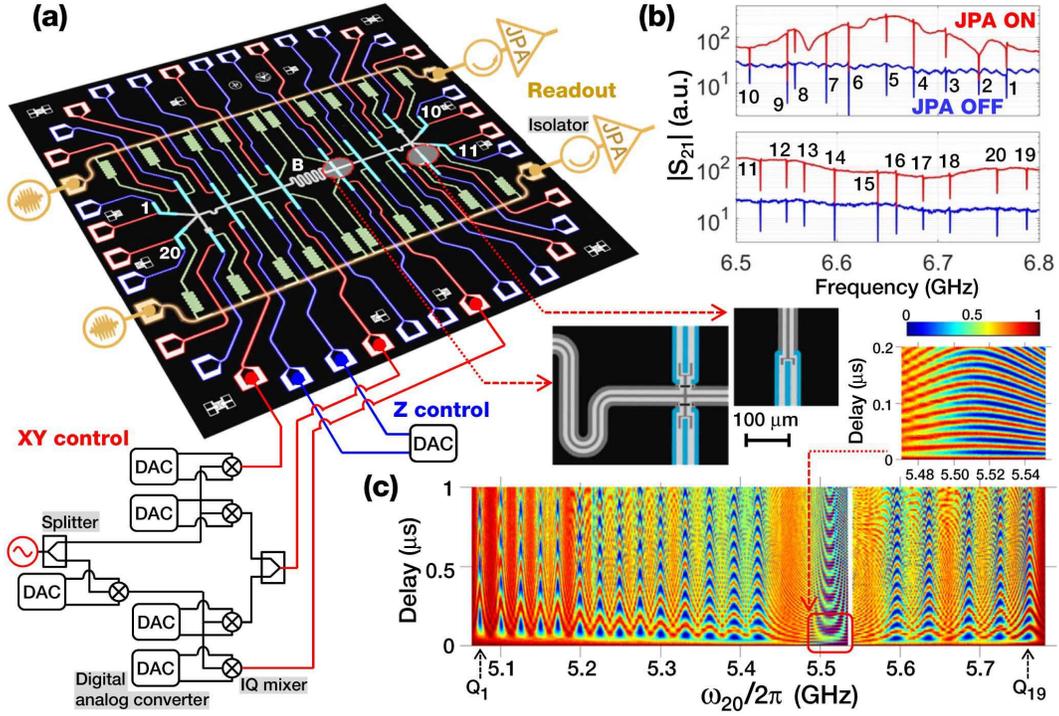}
\caption{ 
(a) False-color circuit image showing 20 superconducting qubits (line shapes in cyan that are labeled clockwise from 1 to 20) 
interconnected by a central bus resonator $B$ (grey). 
Each qubit has its own flux bias line (blue) for Z control, 
and 16 qubits have individual microwave lines (red) for XY control, while
$Q_4$, $Q_7$, $Q_{14}$, and $Q_{17}$ share the microwave lines of neighboring qubits.
Each qubit has its own readout resonator (green), which couples to
one of the two transmission lines (orange) for simultaneous readout.
Also shown are zoomed-in views of representative qubit-bus resonator coupling capacitors 
with different capacitance values at spots as indicated,
and illustrative schematics of the measurement setup showing wirings to the circuit chip, where
we cascade two stages of sideband mixings to generate microwave pulses 
that cover tones with a 1~GHz-bandwidth while maintaining the capability of tracking the relative phases among tones.
(b) Signal spectra through the transmission lines $|S_{21}|$ while all qubits are in $|0\rangle$. Shown are the amplitudes of
the demodulated signals as functions of signal frequency when the JPAs
are ``ON'' (red) and ``OFF'' (blue). The qubits' readout resonators, labeled
from 1 to 20, are visible as dips on the spectra. 
(c) Swap spectroscopy of $Q_{20}$, which is obtained by exciting $Q_{20}$ to $|1\rangle$ and then
measuring its $|1\rangle$-state probability as function of both the qubit frequency and delay time.
The probability data, corrected for elimination of the measurement errors~\cite{Zheng2017}, 
are from two continuous scans separated by the vertical white stripe. 
During the scans the other 19 qubits are sorted in frequency with Z controls and are identified 
by the well-resolved Chevron patterns, which are due to
coherent energy exchanges between $Q_{20}$ and the qubits mediated by the bus resonator $B$.
Zoomed-in view is the direct energy exchange between $Q_{20}$ and $B$. 
}
\label{fig1}
\end{figure*}

In this letter we introduce our latest upgrade, a more powerful 20-qubit superconducting quantum processor featuring
all-to-all connectivity with programmable qubit-qubit couplings mediated 
by a bus resonator. 
With all qubits designed to be uniformly coupled to the bus resonator, 
we engineer a one-axis twisting Hamiltonian by
identically detuning the qubits from the bus resonator. 
Free evolution under the engineered Hamiltonian 
steers the system to squeezed spin states, and then to over-squeezed regime with
suppositions of atomic coherent states at specific time intervals, which are experimentally captured. 
The final GHZ states are characterized by synchronized local manipulations and detections 
of all qubits, and we measure a fidelity figure of 
$0.525\pm 0.005$ for 18 qubits, which confirms the genuine eighteen-partite entanglement~\cite{Guhne2009}.

The new version of the superconducting quantum processor and 
critical peripheral electronics are illustrated in Fig.~\ref{fig1}(a), which consists of
20 frequency-tunable transmon qubits, labeled as $Q_{j}$ for $j$ = 1 to 20, surrounding
a central coplanar waveguide bus resonator ($B$), whose resonant frequency is fixed at 
$\omega_B/2\pi \approx $ 5.51 GHz. Qubit-resonator ($Q_j$-$B$) coupling strengths $g_{j}$ are designed 
to be uniform, and measured $g_{j}/2\pi$ values range from 24.1 to 30.1 MHz. 
Qubits are detected through their respective readout resonators, whose
signal spectra are shown in Fig.~1(b). We use impedance matched 
Josephson parametric amplifiers (JPAs) and an optimized arrangement of the qubit frequencies, $\omega_j^m$, during the readout
to enhance the signal-to-noise ratio.

All qubits are individually tunable with high flexibility, and we show an example
in Fig.~1(c) by measuring $Q_{20}$'s swap spectroscopy while we 
equally space the other 19 qubits in frequency around the resonator $B$. 
Typical qubit energy relaxation times, $T_1$, are in the range of 20 to 50~$\mu$s. 
With a proper arrangement of the qubit idle frequencies, $\omega_j$, where qubit initializations and 
single-qubit rotations are applied, fidelity values of the simultaneous single-qubit $\pi/2$ rotational gates used in the GHZ experiment are 
all above 0.99 as estimated by quantum state tomography and simultaneous randomized benchmarking. 
See Supplemental Material for more details on the device and its operations~\cite{supp}.

\begin{figure}[t]
\includegraphics[clip=True,width=3.0in]{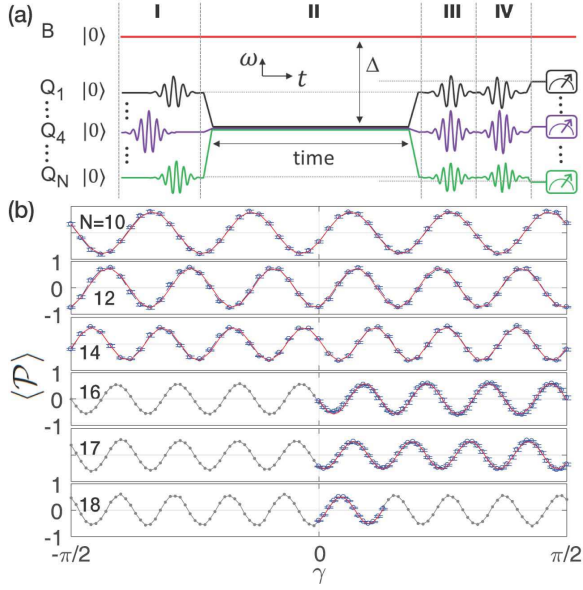}
\caption{
(a) Pulse sequence for generating and characterizing the $N$-qubit GHZ state. 
For a qubit who shares its neighbor's microwave line, e.g., $Q_4$ which shares $Q_5$'s, its first $X_{\pi/2}$
rotational pulse (sinusoids in zone I), which has an amplitude almost ten times larger than 
$Q_5$'s own rotational pulse and therefore may dispersively drive $Q_5$ as well, 
starts earlier while $Q_5$ is in $|0\rangle$ in order to reduce the extra phase picked up by $Q_5$ due to this dispersive drive.
(b) $N$-qubit GHZ parity oscillations. 
For each data point (blue circles), we repeat the state preparation and measurement sequence about $30\times2^N$
times to find the raw $2^N$ occupational probabilities and then apply readout corrections to
eliminate the measurement errors~\cite{Zheng2017},
following which we use maximum likelihood estimation to validate the occupational
probabilities and calculate the parity value $\langle\mathcal{P}\rangle$. 
To estimate error bars, we divide the complete dataset into 
subgroups, each containing about $5\times2^N$ samplings, and the error bars correspond to the standard deviations of those calculated from these subgroups. 
Red lines are sinusoid fits, with the fringe amplitudes corresponding $\rho_{00...0,\, 11...1}$.
For $N = 16$ to 18, repeated measurements with the sampling size of about $30\times2^N$
times over a full range of $\gamma\in[-\pi/2, \pi/2]$ take
too long. Therefore we reduce the range of $\gamma$ that has enough samplings. 
As such grey dots connected by dashed lines are calculated from the experimental data with reduced sampling size, 
which are plotted only for visual guide of the oscillations.
}
\label{fig2}
\end{figure}

With each of the 20 qubits being addressable, the system Hamiltonian 
is
\begin{multline}
H_1 /\hbar = \omega_B a^{\dagger}a + \sum_{j=1}^{20}{\left[\omega_{j}(t)|1_j\rangle \langle 1_j|
+ g_{j}(\sigma _{j}^{+}a+\sigma _{j}^{-}a^{\dagger})\right]} \\
+ \sum_{j=1}^{20} \lambda^\textrm{c}_{j,j+1}(\sigma _{j}^{+} \sigma _{j+1}^{-} +\sigma _{j}^{-}\sigma_{j+1}^{+}),  \label{eq1}
\end{multline}
where $\omega_{j}(t)$ ($\gg g_j$) is tunable within a time scale of a few nanoseconds, 
$\sigma _{j}^{+}$ ($\sigma _{j}^{-}$) is the raising (lowering)
operator of $Q_{j}$, $a^{\dagger}$ ($a$) is the creation (annihilation)
operator of $B$, and $\lambda^\textrm{c}_{j,j+1}$ describes the crosstalk couplings between neighboring qubits
(Subscripts in $\lambda^\textrm{c}_{j,j+1}$ run cyclically from 1 to 20).  
Although more qubits are integrated in this processor, the measured $\lambda^\textrm{c}_{j,j+1}/2\pi$ values
are seen to be reduced from $\sim$2~MHz in the previous 10-qubit version~\cite{Song2017,Xu2018} to around 1 MHz or less 
since we separate the qubits physically as much as possible (see Fig.~1(a)).
Note that there may exist those qubit-qubit crosstalk couplings beyond neighboring pairs, which
should be relatively small and are not included in Eq.~(\ref{eq1}).

\begin{figure*}[t]
\includegraphics[clip=True,width=160mm]{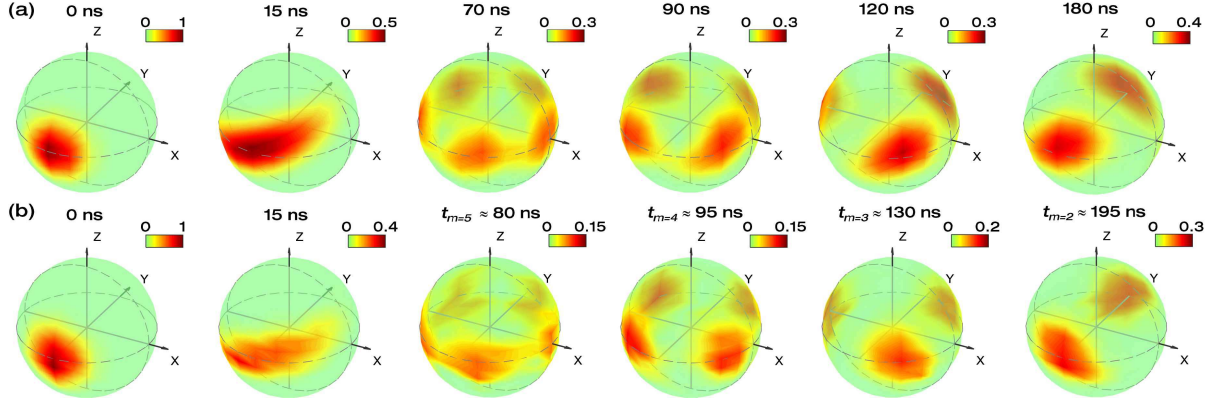}
\caption{Dynamics of the 20-qubit system
illustrated with the quasidistribution $Q$-function. 
(a) Numerical simulations of $Q(\theta,\phi)$ in the spherical polar plots at specific time intervals predicted by Eq.~(1) ignoring decoherence,
after the qubits are initialized in an atomic coherent state $|\pi/2, -\pi/2\rangle$.
(b) Experimental measured $Q_\textrm{exp}(\theta,\phi)$ 
at time intervals as listed. Additional single-qubit dynamical phases, which are accumulated during the qubit frequency tuning process 
and not included in the calculations in (a), are numerically added to rotate the plots in (a) for 
a better visual match with the $Q_\textrm{exp}(\theta,\phi)$ plots in (b). 
The difference in the time steps as listed between (a) and (b) may be due to various factors such as 
uncertainties in some device parameters including $\Delta$ and $g_j$, imperfection in the experimental pulse sequences, 
and, most likely, the existence of the small qubit-qubit crosstalk coupling terms beyond neighboring pairs that are not included in Eq.~(\ref{eq1}).
Note that including all the small crosstalk coupling terms
in the numerical simulation renders the Hamiltonian matrix less sparse and will considerably raise the computational complexity.
}
\label{fig3}
\end{figure*}

As demonstrated previously~\cite{Song2017}, the unique feature of this architecture is that, 
although qubits are physically separated by the bus resonator $B$,
the qubit-qubit coupling mediated by $B$ can be programmed 
with fast Z controls 
to match or detune their frequencies~\cite{Song2017,Guo2018}.
More remarkably, 
in our processor, we can selectively entangle $N$ of the 20 qubits by 
detuning the selected qubits from the resonator by the same amount $\Delta$ ($\gg g_j$),
with the other qubits being far off-resonant. 
When resonator $B$ is initially in vacuum, the effective Hamiltonian for these $N$ qubits, relabeled by $Q_j$ with $j$ going from 1 to $N$,
in the frame rotating at the detuned qubit frequency is~\cite{Zheng2001,Agarwal1997}
\begin{multline}
H_{2}/ \hbar=\sum_{\{j,k\}\in N} {\frac{g_j g_k}{\Delta} \left ( \sigma_{j}^{+}\sigma_{k}^{-} +\sigma_{j}^{-}\sigma_{k}^{+}\right)} 
+ \sum_{j=1}^{N} {\frac{g_j^2}{\Delta} \left| 1_j\right\rangle\left\langle 1_j\right|}\\
+ \sum_{j=1}^{N} {\lambda^\textrm{c}_{j,j+1} \left ( \sigma_{j}^{+}\sigma_{j+1}^{-} +\sigma_{j}^{-}\sigma_{j+1}^{+}\right)},
\label{eq2}
\end{multline}
where $\{j,k\}$ takes all possible pairs within the $N$ qubits and
subscripts in $\lambda^\textrm{c}_{j,j+1}$ run cyclically from 1 to $N$.

The scenario of a system of $N$ identical two-level atoms  
interacting collectively and dispersively with a single mode
electromagnetic field in a cavity has been theoretically investigated~\cite{Agarwal1997,Zheng2001}.
In our experiment we position the qubits 330~MHz below $\omega_B/2\pi$ for all the
effective qubit-qubit couplings (190 terms) in the first summation of Eq.~(\ref{eq2}), $\left|{g_j g_k}/{2\pi\Delta}\right|$, to be $\sim 2$~MHz
while the few ($<20$ terms) neighboring couplings $\lambda^\textrm{c}_{j,j+1}/2\pi$ are from 0.5 to 1~MHz.
Therefore we can ignore $\lambda^\textrm{c}_{j,j+1}$ and those relatively small qubit-qubit crosstalk couplings beyond neighboring pairs (not included in Eq.~(\ref{eq1}))
for now and assume that couplings within all qubit pairs are approximately equal, 
so that the theory predictions~\cite{Agarwal1997,Zheng2001} can be adapted to our experiment.
We emphasize that the imperfection in uniformity has been taken into account by numerical simulations using device parameters based on
the Hamiltonian in Eq.~(\ref{eq1}),
and we find decent agreement between our experimental results and the simplified theoretical treatment in Refs.~\cite{Agarwal1997,Zheng2001}.

With uniform couplings noted as $\lambda = \overline{g_j g_k}/{\Delta}$, 
we now apply the spin representation of qubit states
and define the collective spin operators  
$\mathcal{S}^+ = \sum_j{\sigma_j^+}$, $\mathcal{S}^- = \sum_j{\sigma_j^-}$, and $\mathcal{S}_z = \sum_j{\sigma_{z,j}}$.
The term $\sum{\lambda (\sigma_{j}^{+}\sigma_{k}^{-} +\sigma_{j}^{-}\sigma_{k}^{+})}$ in Eq.~(\ref{eq2})
is then transformed to $\lambda\mathcal{S}^+\mathcal{S}^- \rightarrow -\lambda\mathcal{S}_z^2$ ignoring trivial linear and constant terms,
which is the one-axis twisting Hamiltonian.
By initializing the $N$ qubits identically so that each individual qubit points to the same direction represented by the angles ($\theta$, $\phi$)
in its Bloch sphere, we write down the wavefunction of the atomic (spin) coherent state as 
\begin{equation}
\psi(0) = |\theta, \phi\rangle= 
\left[\cos\frac{\theta}{2}|0\rangle+\sin\frac{\theta}{2} e^{i\phi} |1\rangle\right]^{\otimes N}.
\label{eq.ACS}
\end{equation}
Evolution of the wavefunction under the one-axis twisting Hamiltonian, $H=-\lambda\mathcal{S}_z^2$, was analytically obtained in Ref.~\cite{Agarwal1997}, which shows
that at particular time $t=\pi/m\lambda$, where $m$ is an integer no less than 2,
$\psi(0)$ evolves to a superposition of multiple atomic coherent states, i.e.,
it becomes an atomic Schr\"odinger cat state. 
In particular, at $m=2$, it evolves to a superposition of 
two atomic coherent states, i.e., the $N$-qubit GHZ state,
\begin{multline}
\psi\left(t = \pi/2|\lambda|\right) = e^{-iHt}|\theta, \phi\rangle = \frac{1}{\sqrt{2}}e^{-i(N-0.5)\pi/2} \\ 
\times \left[\left|\theta, \phi - \frac{N-1}{2}\pi\right\rangle + e^{-i\pi/2}\left|\theta, \phi - \frac{N-3}{2}\pi\right\rangle\right].
\label{eq.GHZ}
\end{multline}

Figure~2(a) shows the pulse sequence for generating and characterizing the $N$-qubit GHZ state. We
start with initializing each of the $N$ qubits in $\left(\ket{0}-i\ket{1}\right)/\sqrt{2}$, 
which collectively corresponds to an atomic coherent state $|\pi/2, -\pi/2\rangle$ in the ($\theta$, $\phi$) notation, by applying an $X_{\pi/2}$ rotational pulse
at the qubit's idle frequency (sinusoids in zone I), following which we bias the $N$ qubits
to $\Delta/2\pi \approx -330$~MHz for an optimized duration close to $\pi/2|\lambda|$ (zone II).
The phase of each qubit's XY drive, which defines the rotational axis in the equator plane, is calibrated according to the rotating frame with respect to $\Delta$,
ensuring that all $N$ qubits are in the same initial state just before their collective interactions are switched on~\cite{Song2017,Song2018}.
Right after the interactions we bias these qubits back to their respective idle frequencies, $\omega_j$, for further operations if necessary,
and then to their respective measurement frequencies, $\omega_j^m$, for readout.
We note that during the frequency tuning process qubits may gain different dynamical phases,
i.e., the $x$-$y$ axes rotate differently in the equator planes for different qubits, which
can be determined by a separate phase tracking measurement followed by an optimization procedure (see Supplemental Material~\cite{supp}).

The resulting GHZ state is a superposition of $\left|\pi/2, -N\pi/2\right\rangle$ and 
$\left|\pi/2, -(N-2)\pi/2\right\rangle$ in the collective spin representation,
which can be transformed to a superposition of the $N$ qubits all in $|0\rangle$ and those all in $|1\rangle$
by applying to each qubit a $\pi/2$ rotation around its $x$ ($N$ odd) or $y$ ($N$ even) axis. 
After such a transformation (sinusoids in zone III of Fig. 2(a)), 
the wavefunction is written as $\left(|00...0\rangle +e^{i\varphi}|11...1\rangle\right)/\sqrt{2}$, where $\varphi = \pi/2$ for uniform couplings.
The diagonal elements of the GHZ density matrix $\rho_{00...0}$ and $\rho_{11...1}$ can be directly probed: 
For each state generation and characterization pulse sequence we simultaneously 
measure all qubits which returns an $N$-bit binary string, e.g, 01...0, showing the collapsed multiqubit state;
we repeat the same pulse sequence multiple times and count the probabilities of finding all $N$ bits in 0 for $\rho_{00...0}$ and all those in 1 for $\rho_{11...1}$.

The off-diagonal elements $\rho_{00...0,\,11...1}$ and $\rho_{11...1,\,00...0}$ can be obtained by 
measuring the parity oscillations, defined as the expectation value of the operator 
$\mathcal{P}(\gamma)=\otimes_{j=1}^{N} (\cos{\gamma}Y_j+\sin{\gamma}X_j)$, which is given by
$\langle \mathcal{P}(\gamma) \rangle = 2\left|\rho_{00...0,\, 11...1}\right|\cos(N\gamma+\varphi)$ for the abovementioned GHZ wavefunction~\cite{Monz2011}. 
Experimentally we apply to each qubit a rotation (sinusoids in zone IV of Fig. 2(a)) which bring the axis defined by the operator $\mathcal{P}(\gamma)$, i.e., the direction represented by the angles 
($\pi/2$, $\pi/2-\gamma$) in each qubit's Bloch sphere, to the $z$ axis, followed by simultaneous qubit readout.
Repeating each state generation and measurement pulse sequence multiple times yields $2^N$ probabilities ($P_{00...0}$, $P_{00...1}$, ..., $P_{11...1}$), and
the parity is calculated as $\langle\mathcal{P}\rangle = P_\textrm{even} - P_\textrm{odd}$ with $P_\textrm{even}$ ($P_\textrm{odd}$) corresponding
to the summation of all those probabilities with even (odd) number of qubits in $|1\rangle$.
The clear oscillation patterns of $\langle\mathcal{P}(\gamma)\rangle$, whose amplitude gives $\left|\rho_{00...0,\, 11...1}\right|$, 
confirm the existence of coherence between the two states $|00...0\rangle$ and $|11...1\rangle$ (Fig.~2(b)).
Using values of $\rho_{00...0}$, $\rho_{11...1}$, and $\left|\rho_{00...0,\, 11...1}\right|$ obtained above,
$N$-qubit GHZ state fidelities are calculated to be
$0.817\pm0.009$ ($N=10$), $0.775\pm 0.011$ ($N=12$), $0.655\pm 0.009$ ($N=14$), $0.579\pm 0.007$ ($N=16$), 
$0.549\pm 0.006$ ($N=17$), and $0.525\pm0.005$ ($N=18$), all confirming genuine multipartite entanglement~\cite{Guhne2009}.

Furthermore, detailed dynamics connecting the atomic coherent state in Eq.~(\ref{eq.ACS}) to the GHZ state in Eq.~(\ref{eq.GHZ}) under 
the one-axis twisting Hamiltonian was analytically given in Ref.~\cite{Agarwal1997}, 
where squeezed spin states and more atomic Schr\"odinger cat states other than
the final GHZ state sequentially appear. We are able to take snapshots
of this dynamic process with up to 20 qubits by measuring the quasidistribution $Q$-function
$Q(\theta,\phi) \propto \langle\theta,\phi| \rho(t) |\theta,\phi\rangle$, where $\rho(t)$ is the evolving multiqubit density matrix.
For the 20 qubit case, we bias all qubits to $\Delta/2\pi \approx -470$~MHz since a newly added qubit, $Q_{15}$, is interfered by a two-level state defect at the previous entangling frequency.
To obtain $Q(\theta,\phi)$, we rotate the axis defined by the angles ($\theta$, $\phi$) to
the $z$ axis for each qubit and do so simultaneously for all $N$ qubits before joint readout:
For $\theta<\pi/2$, we rotate by angle $-\theta$ around the $\phi-\pi/2$ axis in the equator plane and record $P_{00...0}$ as $Q_\textrm{exp}$;
for $\theta>\pi/2$, we rotate by angle $\pi-\theta$ around the $\phi+\pi/2$ axis in the equator plane
to reduce the amplitude of the rotational pulse and record $P_{11...1}$ as $Q_\textrm{exp}$.
Obtained values of $Q_\textrm{exp}$ are plotted as functions of $\theta$ and $\phi$ in the spherical polar plots 
as shown in Fig.~3, together with the numerical simulations
ignoring decoherence. We observe the squeezed spin regime at the beginning ($\sim 15$~ns) and the atomic Schr\"odinger cat states 
which are superpositions of $m=5$, 4, 3, and 2 atomic coherent states at $t_m\approx80$, 95, 130, and 195~ns, respectively.
It is seen that $t_m \propto 1/m$ as predicted in Ref.~\cite{Agarwal1997}.
For an $m$-component atomic Schr\"odinger cat state of $N$ qubits, the overlap between adjacent two components 
is $\cos^N(\pi/m)$. Therefore to observe superpositions with more components one needs to increase $N$ to reduce the overlap.
We note that superpositions of up to 4 coherent states have been previously observed in cold atoms and 
superconducting cavities~\cite{Greiner2002,Hofheinz2009,Vlastakis2013}. Here
for the first time we observe the 5-component atomic Schr\"odinger cat state with $N=20$ qubits.
 
In summary, our experiment demonstrates an upgraded and much more powerful version of the multiqubit-resonator-bus
architecture for scalable quantum information processing, with 20 individually
addressable qubits and programmable qubit-qubit couplings. 
Based on this device, we efficiently and deterministically generate the 18-qubit \textit{genuinely entangled} GHZ state and
multi-component atomic Schr\"odinger cat states of up to 20 qubits by
engineering a one-axis twisting Hamiltonian. The high controllability and efficiency of our superconducting quantum processor
demonstrate the great potential of an all-to-all connected circuit architecture for scalable quantum information processing.\\

\noindent{\textbf{Acknowledgments}}.\noindent{
This work was supported by the National Basic Research Program
of China (Grants No. 2017YFA0304300 and No. 2016YFA0300600), the
National Natural Science Foundations of China (Grants No. 11725419 and No. 11434008), and Strategic Priority Research Program of Chinese Academy of Sciences (Grant No. XDB28000000).
Devices were made at the Nanofabrication Facilities at Institute of Physics in Beijing and National
Center for Nanoscience and Technology in Beijing. }

\clearpage
\renewcommand\thefigure{S\arabic{figure}}
\renewcommand\theequation{S\arabic{equation}}
\renewcommand\thetable{S\arabic{table}}

\setcounter{figure}{0}
\setcounter{equation}{0}
\setcounter{table}{0}

\renewcommand\thefigure{S\arabic{figure}}
\renewcommand\theequation{S\arabic{equation}}
\renewcommand\thetable{S\arabic{table}}
\renewcommand\thesection{\arabic{section}}
\renewcommand\thesubsection{\thesection.\arabic{subsection}}

\begin{center}
{\noindent {\bf Supplementary Material for\\
``Observation of multi-component atomic Schr\"odinger cat states of up to 20 qubits''}}
\end{center}

\noindent{\textbf{Qubit-bus resonator coupling.}}
Our sample is a superconducting circuit consisting of 20 Xmon qubits interconnected by
a bus resonator, fabricated with three steps of aluminum depositions: 
Growth of the base wiring layer, double-angle evaporation of the junction bilayer, 
and coating the airbridge layer to connect signal lines and to short grounding pads.
The overall design is similar to that reported in Ref.~\cite{Song2017}.
However, as illustrated in Fig.~1(a) of the main text, a key improvement in the current architecture is that 
the qubits, line shapes in cyan, are far separated as they go around the bus resonator in order
to minimize the XX-type crosstalk couplings between neighboring qubits.
Consequently, some of the qubits have to couple to the bus resonator $B$
at spots away from both ends of the center trace of $B$ via interdigitated capacitors.
For the $Q_j$-$B$ coupling strengths $g_{j}$ to be relatively uniform, we
choose different coupling capacitance values according to the coupling spots, as exemplified in Fig.~\ref{fig1}(a) of the main text.
Measured $g_{j}/2\pi$ values are listed in Tab.~\ref{tab1} and are in agreement with expectations.

\begin{figure}[t]
	\centering
	\includegraphics[width=2.5in,clip=True]{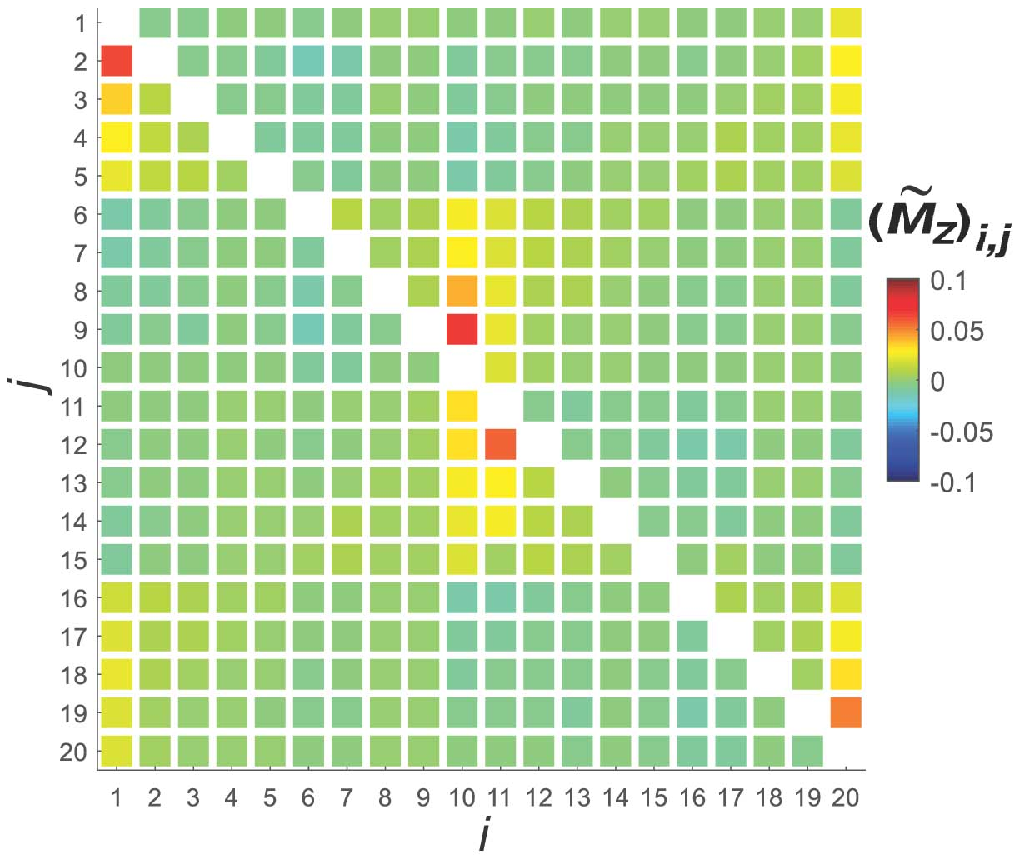}
	\caption{\textbf{Z-crosstalk matrix $\tilde{M}_{Z}$}. Each element $(\tilde{M}_{Z})_{i,j}$ represents the Z bias magnitude sensed by $Q_j$ when $Q_i$ is applied with a unitary Z bias.
		\label{Zcrosstalk}}
\end{figure}

\noindent{\textbf{Qubit manipulation.}}
As shown in Fig.~\ref{fig1}(a) of the main text, each qubit is frequency-tunable via its own flux bias line (Z control in blue) 
and can be coherently driven with its own or neighbor's microwave line (XY control in red).
The effective Z bias of each qubit due to a unitary bias applied to other qubits' Z lines is calibrated, 
which yields the Z-crosstalk matrix $\tilde{M}_Z$ as plotted in Fig.~\ref{Zcrosstalk}. 
The fact that only a few elements of $\tilde{M}_Z$, which describe the Z crosstalk magnitudes between certain neighboring pairs, reach maximum values at about 6.4\%, 
indicates that the airbridges are connecting the grounding pads properly. 

\begin{figure*}
	\centering
	\includegraphics[clip=True,width=145mm]{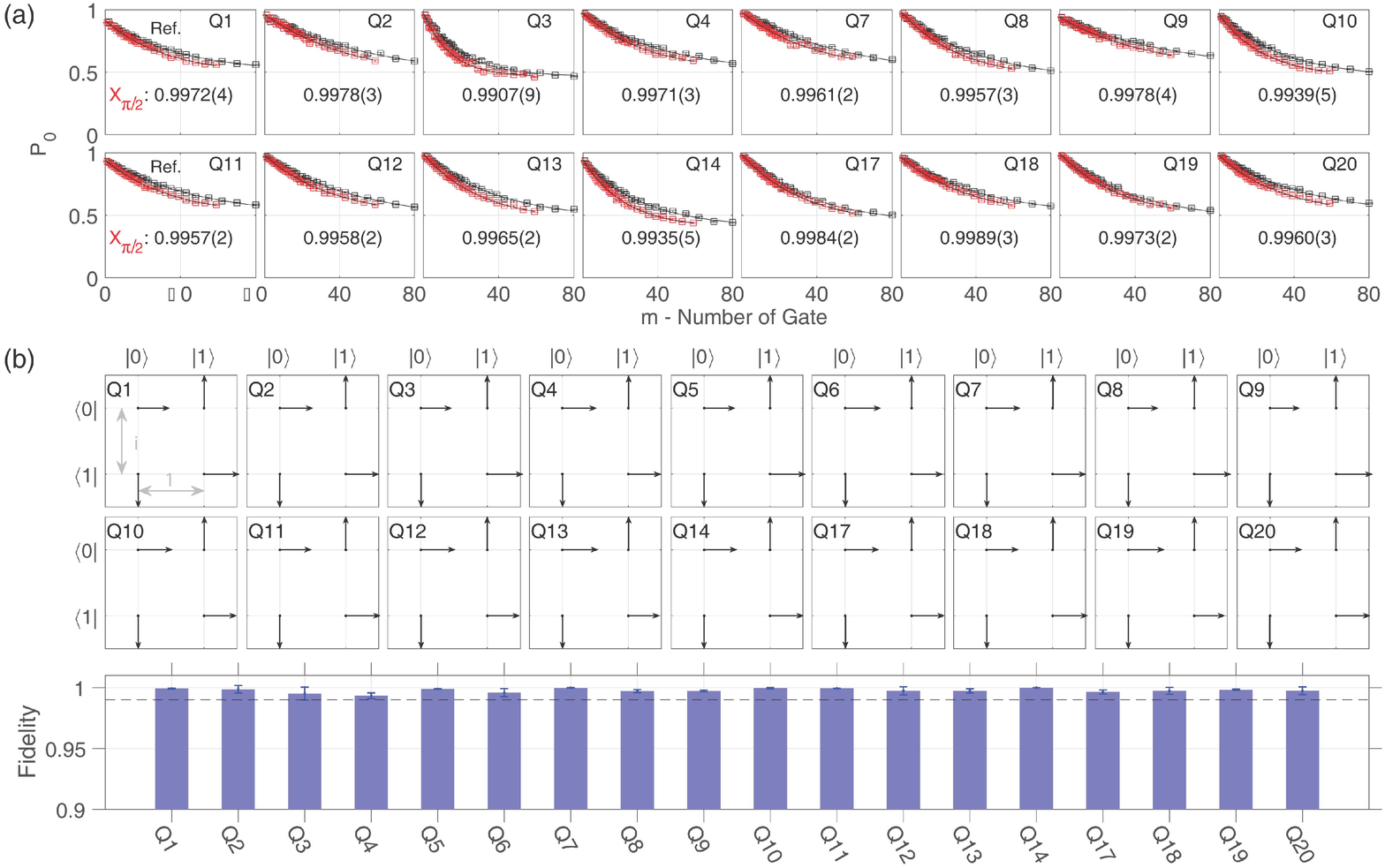}\\
	\caption{\textbf{Calibrations of the single-qubit rotational gates for the GHZ experiment.}
	(a) 16-qubit simultaneous randomized benchmarking (RB) results with the single-qubit $X_{\pi/2}$ gate fidelities as listed. 
	Each pulse sequence of the reference has up to $m$ single-qubit Cliffords for each qubit~\cite{Song2017}. Each Clifford includes additional idling gates to ensure that 
	the $j$-th Cliffords on all 16 qubits can be synchronized. A final Clifford gate returns each qubit to $|0\rangle$.
	At the end of the pulse sequence we measure all 16 qubits simultaneously for $2^{16}$ occupation probabilities, based on which we perform partial trace over other qubits' indices
	for one piece of data on the target qubit; we sum over all data of $k = 30$ random pulse sequences for an exponential fit as shown in each panel.
  On average, each single-qubit Clifford consists of 0.375 gates from $X_{\pi}$, $Y_{\pi}$ and I, 1.5 gates from $\pm X_{\pi/2}$ and $\pm Y_{\pi/2}$, and approximately 1.286 idling 
	gates for synchronization. The pulse lengths of the $X_{\pi}$ and $Y_{\pi}$ gates are 80 ns, and the pulse lengths of all other gates are 40 ns.
	Black dots are the $|0\rangle$-state probability data, $P_0$, of each qubit as function of $m$ for the reference, 
	and red dots are data with $X_{\pi/2}$ inserted.
	(b) Single-qubit quantum state tomography characterizing the state $\left(|0\rangle-i|1\rangle\right)^{\otimes 18}$ prepared by applying $\pi/2$ gates on the 18 qubits,
	with the state preparation pulse sequence listed in zone I of Fig.~2(a) in the main text. 18 single-qubit density matrices are shown in the upper row, 
	where the amplitude and phase of a matrix element are represented by the length and direction, respectively, of an arrow in the complex plane. 
	The state fidelity metrics, shown in the bottom row, are all above 0.990 which agree with the RB results in (a).}
	\label{rb_qst}
\end{figure*}

Taking into account the qubit's weak anharmonicity, we carefully arrange the resonant frequencies of all qubits, $\omega_j$, where qubits
are initialized and operated with single-qubit rotational gates, to minimize any possible crosstalk errors, e.g., unwanted ZZ-type couplings. 
As such $\omega_j/2\pi$ values have to spread out as much as possible 
in a frequency span from 4.3 to 5.3~GHz (see $\omega_j$ in Tab.~\ref{tab1}).
To cover such a wide range of microwave tones while maintaining the capability of tracking
the relative phases among tones, we cascade two stages of sideband mixings as illustrated in Fig.~\ref{fig1}(a) of the main text. 
All together, our custom digital-to-analog converters, through the IQ mixing with a single-tone continuous microwave,
can output phase-tracked microwave pulses with up to 20 tones targeting all 20 qubits for simultaneous XY rotational gates.
Both quantum state tomography and simultaneous randomized benchmarking 
indicate that the gate fidelity values of the 40 ns-long $\pi/2$ rotations, used for the GHZ experiment, are above 0.99 (Fig.~\ref{rb_qst}).

\begin{figure}[!htb]
	\centering
	\includegraphics[width=2.8in,clip=True]{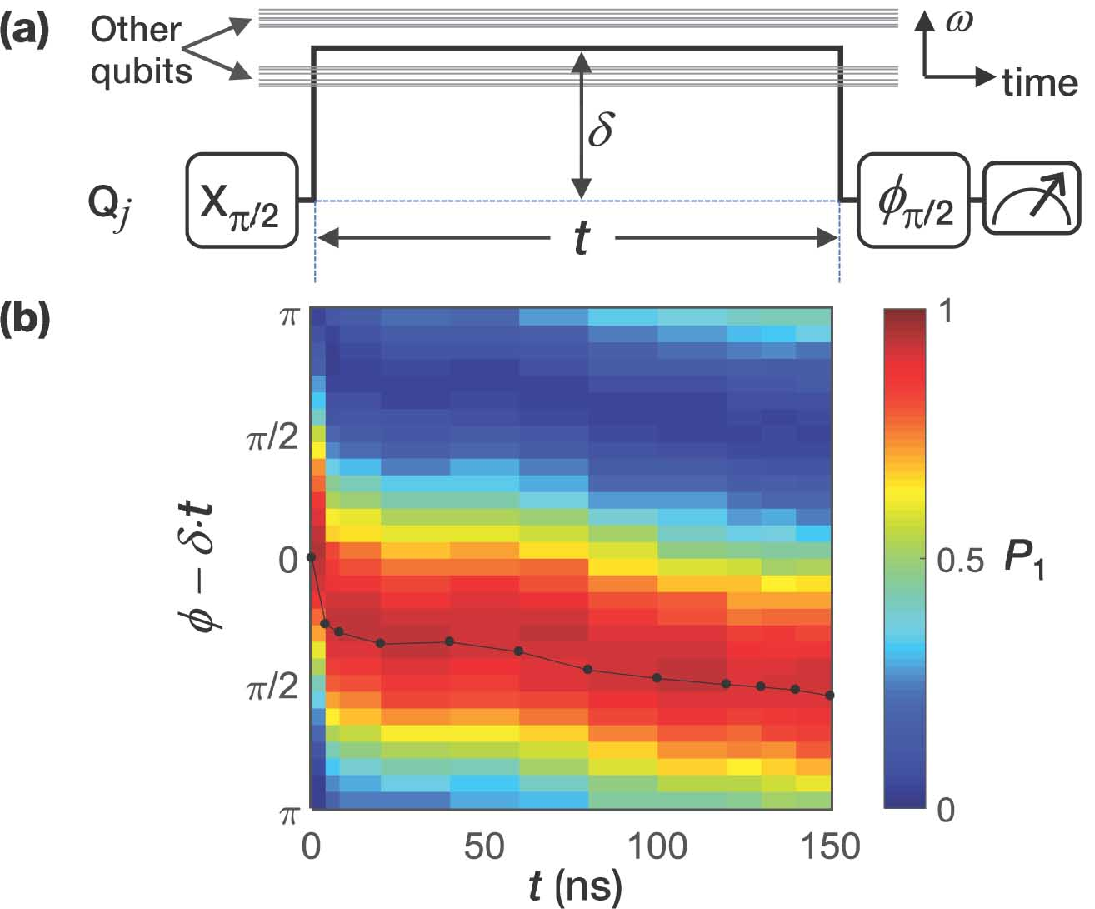}
	\caption{\textbf{Phase tracking measurement.} (a) Pulse sequence of the Ramsey interference measurement used to determine the orientation
	of the $x$-$y$ axes in the Bloch sphere for $Q_j$. 
	(b) Representative experimental data with the pulse sequence shown in (a).
	\label{phaseCal}}
\end{figure}

\noindent{\textbf{Qubit readout.}}
As shown in Fig.~\ref{fig1}(a) of the main text, each qubit, $Q_j$, dispersively interacts with its own readout resonator 
(green) that has a fixed tone at $\omega^r_j$. All readout resonators couple 
to one of the two transmission lines (orange) across the circuit chip. 
At the output of each transmission line, an impedance matched Josephson parametric amplifier (JPA)
is used to enhance the signal-to-noise ratio. To analyze the state of $Q_j$, we
first quickly tune $Q_j$'s frequency to $\omega^{m}_{j}$ via its Z control, and then pump its
readout resonator to be populated with dozens of photons using microwave pulse through the corresponding
transmission line. Since the qubit eigenstates $|0\rangle$ and $|1\rangle$ in the Pauli $Z$ basis 
(along the $z$ axis of the Bloch sphere) affect its readout resonator differently, 
we can observe the transmitted $S_{21}$ signal near $\omega^r_j$ to tell the state of the qubit.
Passing multi-tone signals targeting all $\omega^r_j$ through the transmission lines allows us to simultaneously 
read out $N$ qubits by demodulating the $S_{21}$ signals near their $\omega^r_j$ with custom analog-to-digital converters,
which differentiates the 0 and 1 outcomes for each qubit and returns a joint outcome described 
by an $N$-bit binary string, e.g., $01...0$, showing the collapsed state of $N$ qubits.
We run the same experimental sequence about $30 \times 2^N$ times in order to obtain reliable raw probabilities,
\{$P_{00...0}$, $P_{00...1}$, ...., $P_{11...1}$\}, for $2^N$ basis states.
Measurement in the $X$ ($Y$) basis is achieved by inserting a Pauli $Y$ ($X$) rotation on each qubit.
The qubit $|0\rangle$ and $|1\rangle$-state readout fidelity values, $F_{0,j}$ and $F_{1,j}$, are summarized in Tab.~\ref{tab1},
which are used to correct the raw probabilities to eliminate the readout errors as done previously~\cite{Zheng2017}. 
It is noted that for simultaneous readout, since qubits move down in frequency due to
their readout resonators being populated with microwave photons, $\omega^m_j$ have to 
be carefully arranged to minimize any possible pairwise crosstalk influence (see $\omega^{m}_{j}$ in Tab.~\ref{tab1}).

\noindent{\textbf{Phase tracking measurement.}}
To perform the measurement of the quasidistribution $Q$-function, $Q(\theta,\phi)$, we need to identify the orientation 
of the $x$-$y$ axes in the Bloch sphere for each qubit. In the experiment, 
as qubits are tuned to the interacting frequency, $\omega_\textrm{I}$, for entanglement 
and then back to idle frequencies for further operations with rectangular pulses (see Fig.~\ref{phaseCal}(a)), dynamic phases are accumulated 
which rotate the $x$-$y$ axes in the Bloch sphere for each qubit by an angle proportional to the interacting time $t$. 
Precise calibration of these dynamic phases are needed for the following single-qubit rotational operations 
to ensure the reliability of the $Q$-function measurement. 

Figure~\ref{phaseCal} shows how we track the dynamic phases in this process. We use a Ramsey interference measurement
with the representative pulse sequence shown in Fig.~\ref{phaseCal}(a) to calibrate one qubit at a time. Following the first $\pi/2$ rotation pulse, $X_{\pi/2}$,
and after the target qubit $Q_j$ returns to its idle frequency,  
a second $\pi/2$ pulse whose rotation axis is of phase $\phi$ in the $x$-$y$ plane, $\phi_{\pi/2}$, 
is applied before the qubit is read out. To minimize the $Q_j$'s frequency shift due to all other qubits' existence through the ZZ-type crosstalk, we
distribute the other qubits in the vicinity of the interacting frequency $\omega_\textrm{I}$, which is about 30 to 70~MHz away from $\omega_\textrm{I}/2\pi$, 
so that their combined effect on $Q_j$ is the same as that when all qubits are biased to $\omega_\textrm{I}$ as calculated 
by the measured Z-crosstalk matrix $\tilde{M}_{Z}$ (Fig.~\ref{Zcrosstalk}).
The representative $|1\rangle$-state probability, $P_1$, data as functions of both the interacting time $t$ and the phase difference $\phi-\delta\cdot t$,
where $\delta = \omega_{j} - \omega_\textrm{I}$ is the qubit frequency detuning,
are plotted in Fig.~\ref{phaseCal}(b). 
By tracing the $P_1$ maximum values along the sliced data at each $t$ value, shown as black dots in Fig.~\ref{phaseCal}(b),
we can estimate the extra rotated angle of the $x$-$y$ axes that may be due to the imperfect experimental rectangular pulse.
With the new orientation of the $x$-$y$ axes in the Bloch sphere for this qubit being roughly located, we then perform
an optimization search to fine tune the extra rotated angles of all qubits to maximize the fidelity of the experimentally detected state.

\begin{table*}[b]
	\centering
	\begin{ruledtabular}
	\begin{tabular}{cccccccccccccc}
		&$\omega_{j}^0/2\pi$&$\omega_{j}/2\pi$&$T_{1,j}$&$T_{2,j}^*$&$\lambda_{j,j+1}^c/2\pi$&$g_j/2\pi$&$\omega_j^r/2\pi$&$\omega_{j}^{m}/2\pi$&$F_{0,j}$&$F_{1,j}$\\
		&              (GHz)&            (GHz)& ($\mu$s)&   ($\mu$s)&              (MHz)&    (MHz) &            (GHz)&                (GHz)&         &         \\
		\hline
		$Q_1$    &5.698&4.320&$\sim$23 &2.0 &0.75 &27.6&6.768&4.510 & 0.929&0.887\\
		$Q_2$    &5.611&4.791&$\sim$27 &2.4 &0.83 &27.4&6.741&4.794 & 0.969&0.925\\
		$Q_3$    &5.793&5.330&$\sim$26 &2.0 &1.01 &29.1&6.707&5.295 & 0.973&0.920\\
		$Q_4$    &5.729&4.865&$\sim$35 &1.8 &1.02 &27.6&6.676&4.491 & 0.941&0.922\\
		$Q_5$    &5.585&4.490&$\sim$30 &2.5 &-0.39 &26.5&6.649&4.435 & 0.946&0.911\\
		$Q_6$    &5.450&4.350&$\sim$29 &3.0 &1.07 &29.2&6.611&4.310 (4.300) & 0.927&0.893\\
		$Q_7$    &5.480&4.830&$\sim$36 &2.7 &1.10 &27.8&6.589&4.399 & 0.967&0.885\\
		$Q_8$    &5.560&4.965&$\sim$37 &2.5 &0.83 &30.1&6.558&4.905& 0.954& 0.919\\
		$Q_9$    &5.583&4.290&$\sim$20 &2.9 &0.79 &24.1&6.551&4.370 & 0.933&0.896\\
		$Q_{10}$ &5.583&5.290&$\sim$33 &2.7 &0.65 &27.7&6.513&5.375 (5.345) & 0.977 (0.967)&0.846 (0.908)\\
		$Q_{11}$ &5.682&4.425&$\sim$35 &2.8 &0.77 &27.3&6.524&4.290 (4.340) & 0.943&0.889\\
		$Q_{12}$ &5.690&5.250&$\sim$33 &1.8 &0.81 &26.9&6.550&5.345 (5.375) & 0.981 (0.977)&0.876 (0.903)\\
		$Q_{13}$ &5.660&4.899&$\sim$31 &2.0 &0.96 &29.1&6.568&4.819 & 0.986&0.934\\
		$Q_{14}$ &5.723&5.220&$\sim$51 &2.4 &1.08 &27.4&6.598&4.885 & 0.993&0.951\\
		$Q_{15}$ &$\sim$5.7&4.290&$\sim$24 &2.1&-0.21 &26.3&6.640&4.34 &0.981 &0.903\\
		$Q_{16}$ &5.642&4.260&$\sim$37 &2.8&0.77 &26.5&6.659&4.01 &0.966 &0.925\\
		$Q_{17}$ &5.843&4.700&$\sim$51 &2.3 &0.91 &27.3&6.685&4.850 & 0.989&0.940\\
		$Q_{18}$ &5.775&4.385&$\sim$37 &1.2 &0.54 &29.0&6.712&4.465 & 0.967&0.924\\
		$Q_{19}$ &5.793&5.170&$\sim$46 &2.0 &0.67 &24.6&6.788&4.930 & 0.950 (0.988)&0.875 (0.914)\\
		$Q_{20}$ &5.847&4.766&$\sim$37 &1.7 &0.64 &27.5&6.758&5.839 & 0.991&0.813\\
	\end{tabular}
	\end{ruledtabular}
	\caption{\label{table1} \textbf{Detailed device parameters and qubit performance metrics.} $\omega_j^0$ is $Q_j$'s maximum resonant frequency at the sweet point, 
	and $\omega_{j}$ is $Q_j$'s idle frequency where $Q_j$ is initialized and single-qubit rotational pulses are applied. 
	$T_{1,j}$ and $T_{2,j}^*$ are the typical single-qubit energy relaxation time 
	and Ramsey dephasing time (Gaussian decay)~\cite{Song2017}, respectively, 
	which are estimated based on the data measured around the interaction frequency 
	where multiqubit GHZ states are generated. Note that $T_{1,j}$ may fluctuate over time due
	to the presence of two-level state defects~\cite{Klimov2018} and $T_{2,j}^*$ may not
	be relevant to the experiment since the qubits coupled with each other at the interacting frequency act as a coherent system which
	may become insensitive to the local flux noise within individual qubits~\cite{Xu2018}.
	$\lambda_{j,j+1}^c$ describes the crosstalk coupling strength between neighboring qubits, where $j$ runs cyclically from 1 to 20. 
	$g_j$ is the coupling strength between $Q_j$ and the bus resonator $B$. 
	$\omega_{j}^{r}$ is the resonant frequency of the readout resonator for $Q_j$.
	$\omega_{j}^{m}$ is the resonant frequency of $Q_j$ at the beginning of the qubit readout when $Q_j$'s readout resonator is unpopulated.
	Those values of $\omega_{j}^{m}$ for $Q_6$, $Q_{10}$, $Q_{11}$, and $Q_{12}$ in parentheses are used for 
	the $N$-qubit GHZ experiment with $N$ up to 18, during which
	$Q_{15}$ and $Q_{16}$ are biased to $\sim$4 GHz and can be ignored.
	$F_{0,j}$ ($F_{1,j}$) is the typical probability of detecting $Q_j$ in $\vert 0\rangle$ ($\vert 1\rangle$) 
	when it is prepared in $\vert 0\rangle$ ($\vert 1\rangle$), which is used to correct raw probability data 
	for elimination of the measurement errors~\cite{Zheng2017}. Values in parentheses are for the GHZ experiment:
	As we switched from the 18-qubit GHZ experiment to the 20-qubit atomic Schr\"odinger cat state experiment, $Q_{15}$ and $Q_{16}$ were included
	so that we reconfigured the $\omega_{j}^{m}$ values of 4 qubits, and noticeable drops of a few percent 
	in $F_{1}$ for $Q_{10}$, $Q_{12}$, and $Q_{19}$ were captured since we adjusted the working parameters of Josephson parametric amplifiers. 
	}
	\label{tab1}
\end{table*}

\end{document}